\documentclass[11pt,a4paper,notitlepage]{article}

\usepackage{amsmath}        
\usepackage{amsfonts}       
\usepackage{amsthm}         
\usepackage{bbding}         
\usepackage{bm}             
\usepackage{graphicx}       
\usepackage[square,sort,comma,numbers]{natbib}         
\usepackage[nottoc]{tocbibind} 
\usepackage{paralist}       
\usepackage[usenames]{xcolor}  

\usepackage{xfrac}
\usepackage{tikz}
\usepackage{tikz-cd}

\usetikzlibrary{arrows,shapes,positioning}
\usetikzlibrary{decorations.markings}

\usetikzlibrary{arrows, matrix}

\usepackage{wrapfig}
\usepackage{caption}
\usepackage{float}

\usepackage{titlesec}
\usepackage{filecontents}
\usepackage{mathtools}

\usepackage{mdwlist}
\usepackage{mathrsfs}
\usepackage[shortlabels]{enumitem}
\usepackage{amssymb}
\usepackage{textcomp}
\usepackage[toc,page]{appendix}

\usepackage{pgfplots}
\pgfplotsset{compat=1.7}

\usepackage[utf8]{inputenc} 
\usepackage{authblk} 

\usepackage[cal=boondox,calscaled=1.0]{mathalfa}

\usepackage{physics}

\numberwithin{equation}{section}
\numberwithin{figure}{section}
\theoremstyle{definition}

\usepackage{natbib}
\usepackage[nottoc]{tocbibind}

\title{Discrete Black Holes}
\author[1]{Jakub Káninský}

\affil[1]{Charles University, Faculty of Mathematics and Physics, Institute of Theoretical Physics. E-mail address: jakubkaninsky@seznam.cz}

\date{\today}
\titleformat{\section}{\normalfont\scshape\large}{\thesection}{2em}{}
\titleformat{\subsection}{\normalfont\scshape\normalsize}{\thesubsection}{1em}{}
\titleformat{\subsubsection}{\itshape\normalsize}{\thesubsubsection}{1em}{}

\renewenvironment{abstract}
 {\small
  \begin{center}
  \end{center}
  \list{}{
    \setlength{\leftmargin}{15mm}%
    \setlength{\rightmargin}{\leftmargin}%
  }
  \item\relax}
 {\endlist}

\usepackage{tocloft}
\begin{document}

\maketitle

\begin{abstract}
The classical spacetime is usually described by a differentiable manifold with infinitely many degrees of freedom. Occasionally though, it is useful to consider an approximation whose number of degrees of freedom is finite. There are several discrete models of spacetime like that, some of which have been used to build a (simplified) representation of a black hole. We will shortly revisit these discrete black hole models. Then we limit ourselves to one particular case and show how it can be inhabited by quantum matter fields. It is suggested that the field dynamics should be described by the framework of discrete canonical evolution, and we point out some of the most significant implications of this approach.
\end{abstract}

\vspace{3\baselineskip}

\section{Introduction}

Overall, there are three major reasons for considering a discrete model of spacetime. The first reason is the employment of numerical simulations which often operate with some kind of spacetime lattice in order to find approximate solutions of the Einstein equations. Among other things, these have been used for extensive investigations of black holes, from the two-body systems which are prominent sources of gravitational waves and have recently attracted great attention due to the rise of interferometry \cite{Buonanno2014} to various configurations of three-body systems \cite{Lousto2008, Galaviz2010} and much more \cite{Sperhake2013, Okawa2014, Fendt2019}. Numerical computations based on lattices also play an important role in cosmology. Here, the link to black holes is present in the approach of \textit{black hole lattices} \cite{Bentivegna2018, Gregoris2020} allowing to build large scale cosmological models from sets of discrete matter sources.

The second reason for going discrete is to study the expected effect of quantum gravity which is commonly presumed to introduce some nonzero minimal length-scale, probably of Planckian order \cite{Garay1994, Hossenfelder2013, Padmanabhan2015, Hooft2016}. In some way or another, this aspect is present in the majority of approaches to quantum gravity like the string theory \cite{Szabo2002}, loop quantum gravity \cite{Gambini2011}, causal set theory \cite{Henson2010}, and others. Meanwhile, it is argued that the (built-in or effective) minimal length-scale of spacetime would have deep theoretical and phenomenological implications within areas like quantum field theory, high energy physics, black hole physics and cosmology \cite{Mead1966}. This motivates a large volume of research based on various effective models of discrete spacetime inspired by quantum gravity which attempt to study the (potentially observable) effects of the minimal length-scale. In context of black holes, these may include the backreaction on gravity \cite{Maziashvili2013}, bounds on the creation of very small black holes \cite{Ali2012}, serious consequences for certain physical processes like proton decay \cite{Bambi2008} or effects on the Hawking radiation \cite{Corley1998}, black hole entropy \cite{Mejrhit2019}, information loss \cite{Lowe2015} and the central singularity \cite{Spallucci2011}. There is also a noteworthy line of research exploring the analogy of quantum black holes and the Hydrogen atom \cite{Bekenstein1999, Corda2020}.

The third reason for establishing discrete models of spacetime is their direct employment in certain approaches to quantum gravity. They are central to a handful of significant non-perturbative path-integral approaches like causal sets \cite{Henson2010, Surya2019}, quantum Regge calculus \cite{Hamber2009, Barrett2018} or causal dynamical triangulations \cite{Loll2019} just to name a few. The main advantage of the discrete system over the continuous one is the relative simplicity of its phase space: it only takes a finite number of degrees of freedom to describe a given compact region of spacetime, which opens the door to standard quantum-mechanical construction of the Hilbert space and the representation of the canonical commutation relations. It also allows for the employment of some kind of \textit{sum over histories}. The respective models of spacetime used in these approaches can be very different, as illustrated by the examples mentioned above, and relating them to their continuous analogue is not always straightforward: they become important subjects of research themselves. Besides studies of the general properties, one can also encounter papers that provide discrete analogues of known continuous solutions of the Einstein equations. They often feature black hole spacetimes which have been studied within causal sets \cite{Dou2003, Rideout2006, He2008, Asato2019, Surya2020}, Regge calculus \cite{Wong1971, Khatsymovsky2020} or causal dynamical triangulations \cite{Dittrich2006}. They will be briefly revisited in the next two sections.

Besides pure gravity, discrete spacetime models listed in the previous paragraph can be useful for modeling quantum fields on curved backgrounds. Of course, it is expected that these should be ultimately coupled to gravity, giving rise to a full unified quantum theory. As of now though, such theory is far out of reach. Nevertheless, establishing quantum field theory on discrete fixed background is very much possible as well as relevant, since it provides a viable discrete analogy of the quantum field theory on classical curved spacetime. The problem has been addressed in many works, most often for Regge lattices \cite{Sorkin1975, Hamber1993, Hamber2009, McDonald2010a, Paunkovic2016} but more recently also for causal sets \cite{Sverdlov2008, Johnston2008, Johnston2010, DableHeath2020}. This allows one to study quantum fields non-perturbatively in a number of interesting cosmological or phenomenological scenarios. The aim of the present letter is to advocate for an application to the case of black holes, which are---as we have illustrated---of great interest for a number of reasons. In the last section, we shall outline one particular framework which can be used to one's advantage.

\section{Causal Sets}
Causal set \cite{Surya2019} is a partially ordered locally finite set, i.e., a set $ C $ with relation $ \preceq $ satisfying acyclicity ($ x \preceq y $ and $ y \preceq x ~ \Rightarrow ~ x = y $ for all $ x,y \in C $), transitivity ($ x \preceq y $ and $ y \preceq z ~ \Rightarrow ~ x \preceq z $ for all $ x,y \in C $) and local finiteness (for every $ x, y \in C $, the order interval $ I(x,y) = \{ w \in C ~ \vert ~ x \preceq w \preceq y, ~ x \neq w \neq y \} $ between $ x $ and $ y $ is finite). The elements of $ C $ represent building blocks of the discrete spacetime and the relation $ \preceq $ encodes their causal structure. The volume element needed to fix the residual conformal freedom is determined by counting the elements of $ C $ inside a given spacetime region.

Casual set can be generated from the continuous manifold by a process called \textit{sprinkling}, in which one selects points of the manifold uniformly at random (the average number of selected points in a region is made proportional to the region's volume) and imposes partial ordering induced by the causal order of the points. In this way, one can relatively easily build a causal set model of any classical solution, including black hole spacetimes. For the Schwarzschild solution, this was done in \cite{He2008}. The resulting set has the expected causal structure of the event horizon (no causal links pointing from the interior to the exterior). In the vicinity of the central singularity, it is less ordered due to the narrowing lightcones.

Since local finiteness only restricts the number of elements inside the order interval, the causal set as defined above may include infinite or even uncountable subsets. For instance, it may contain \textit{singular antichains}, i.e., subsets which are uncountable and totally unordered. In \cite{Asato2019}, it is argued that these may correspond to singularities of the continuous theory. Based on this observation, the author suggests the following definition of a causal set black hole: for a singular antichain $ S $, one defines the associated black hole as $ B_{S} = \{ x \in C ~ \vert ~ p^{+}(x) \cap S \neq \varnothing \text{ for all } p^{+}(x) \} $ where $ p^{+}(x) $ is an inextensible future-oriented causal path beginning at $ x $. In other words, if every future-oriented causal path beginning at $ x $ intersects the antichain $ S $, then $ x $ lies inside the black hole. In yet another words, one could say that $ B_{S} = D^{-}(S) $ is the past domain of dependence of $ S $ \cite{Joshi1993}. It appears that singular antichains cannot be actually obtained by sprinkling so their role as causal set singularities is limited; however, the latter definition can be used for an arbitrary set $ S $ representing the singularity.

The partial ordering of a causal set $ C $ is fully determined by its unique collection of \textit{links}, i.e., the relations between couples of distinct elements $ x, y \in C $ ($ x \neq y $) such that $ x \preceq y $ and $ I(x,y) = \varnothing $. Given all the links of $ C $, its causal order is easily induced via transitivity. Links can be therefore viewed as representing the irreducible relations. In \cite{Dou2003}, the number of links between two regions of a causal set was used as a measure of their mutual entanglement, determining the so-called \textit{entanglement entropy}. It was then argued that the black hole entropy can be related to the number of links spanning across the event horizon of a causal set black hole. For several different cases, the entropy was indirectly shown to be proportional to the horizon area.

As noted in the introduction, there are works addressing the issue of inhabiting causal sets with quantum fields \cite{Sverdlov2008, Johnston2008, Johnston2010, DableHeath2020}. However, it will presumably take more time before we can see some real-world applications like the study of quantum field dynamics in the proximity of a causal set black hole.

Causal sets are very economical in the sense that they contain very small---if not minimal---amount of information needed to provide an approximate discrete representation of the continuous Lorentzian manifold. An unpleasant consequence of this principle is that the (approximate) reconstruction of geometry from the causal set is a nontrivial task \cite{Eichhorn2019}. For this reason, causal set is usually not the model of first choice when it comes to practical computations within discrete spacetimes. The most common alternative and one its remarkable spin-off are described in the next section.

\section{Regge Triangulations}

In 1961, it was suggested by Regge \cite{Regge1961} that a continuous differentiable manifold of dimension $ d $ could be approximated by a piecewise-flat one by gluing flat $ d $-dimensional blocks along their faces. The designated building blocks of the piecewise-flat space are \textit{$ d $-simplices}. An $ n $-simplex $ \sigma^{n} $ in $ \mathbb{R}^{d} $ ($ d \geq n  $) is the convex span of $ n+1 $ affinely independent points in $ \mathbb{R}^{d} $ called \textit{vertices} \cite{Ambjorn2005}. (In the Lorentzian case, just replace $ \mathbb{R}^{d} $ with the Minkowski space $ \mathbb{M}^{d} $.) A 0-simplex is a point, 1-simplex is a line segment, 2-simplex is a triangle, 3-simplex is a tetrahedron, and so on. Every $ n $-simplex contains $ {n+1}\choose{k+1} $ $ k- $simplices in its boundary, in particular, it has $ \frac{n(n+1)}{2} $ edges ($ k = 1 $). The values of the edge lengths uniquely specify the geometry. The simplex spanned by a subset of vertices of $ \sigma^{n} $ is a \textit{subsimplex} of $ \sigma^{n} $. A $ (d-1) $-dimensional subsimplex of a $ d $-simplex in $ \mathbb{R}^{d} $ is called a \textit{face}, a $ (d-2) $-dimensional subsimplex is called a \textit{hinge}.

Regge calculus soon turned out to be a powerful tool for studying discretized manifolds and their geometrical properties. It comes with well known expressions for volumes and curvatures corresponding to their continuous analogues \cite{Cheeger1984} which allows for the introduction of Einstein-Hilbert action and the discrete version of Einstein equations \cite{Hamber2009}. The clear analogy between piecewise linear and continuous manifolds makes Regge triangulations well suited for practical computations. For instance, it can be applied to obtain discrete versions of exact black hole solutions like the Schwarzschild or Reissner-Nordstr\"{o}m geometries \cite{Wong1971, Khatsymovsky2020}.

Over the years, Regge calculus has lead to several established non-perturbative approaches to quantum gravity \cite{Oriti2009}. This is the case of \textit{causal dynamical triangulations} \cite{Loll2019} which implement two further assumptions about the triangulation. First, it is assumed that all the vertices belong into individual non-intersecting time-slices labeled by a discrete time parameter and that the slice topology is fixed (this is motivated by a causality argument). Second, the triangulation is assumed to be composed of standardized building blocks, which is helpful for fixing gauge freedom \cite{Romer1986}. In $ d $ dimensions, these building blocks are $ d $-simplices whose spacelike edges have squared edge length $ a^{2} > 0 $ and timelike edges have squared edge length $ - \alpha a^{2} $ where $ \alpha $ is a positive real constant \cite{Ambjoern2013}. Edges between vertices which belong to the same time-slice are always spacelike; edges between vertices which inhabit neighboring slices are always timelike. Note that there are no null edges. All the time-slices are equidistant, as if marked by a uniformly ticking (proper) time parameter. The standardized building blocks come in several types. In dimension $ d = 2 $ there are only two types of triangles: (1,2) which has one vertex at the sooner time-slice and the remaining two on the later one, and (2,1) whose configuration is the opposite. In dimension $ d = 3 $ the triangulation is made of three types of tetrahedra, namely (1,3), (2,2) and (3,1). Analogically, for $ d = 4 $ the triangulation consists of 4-simplices which come in four types. In general there are $ d $ types. Because the building blocks are standardized, the geometry of the resulting simplicial spacetime depends only on the occurrence of these types.

The above brings us to the work \cite{Dittrich2006} in which the authors present a causal dynamical triangulation corresponding to a black hole spacetime, in particular the Kottler solution with the line element
\begin{equation}\label{kottler}
	ds^{2} = - \left( 1-\frac{2M}{r}-\frac{r^{2}}{L^{2}} \right) dt^{2} + \left( 1-\frac{2M}{r}-\frac{r^{2}}{L^{2}} \right)^{-1} dr^{2} + r^{2} d\Omega^{2}
\end{equation}
which was chosen because it conforms to the assumptions of the theory. They also build a horizon finder based on counting distinct types of building blocks, and introduce the concept of \textit{Lorentzian triangulations of product type} whose utility goes beyond the application at hand. The causal dynamical triangulation example is worth attention especially because it is considerably simplified by the use of standardized $ d $-simplices. This allows for measuring various geometrical quantities (e.g. volume or curvature) by means of counting certain structures present within the triangulation---a feature which reminds of causal sets. The strategy of assembling complex geometries from a handful of building blocks can be helpful even outside the strict scope of causal dynamical triangulations; for example, upon relaxing some of the assumptions and taking advantage of the existing tools in order to study phenomenological or other questions. We have seen that in connection to black holes, such questions are abundant.

Thanks to their direct geometrical interpretation, Regge lattices can be enhanced with matter or gauge fields without greater difficulties. The interested reader may find various works providing prescriptions for the lattice field action \cite{Sorkin1975, Hamber2009, McDonald2010a}. For example, the discrete analogue of the classical scalar field action
\begin{equation}\label{scont}
	S = \frac{1}{2} \int_{\Omega} \left( \nabla_a \varphi \nabla^{a} \varphi + m^{2} \varphi^{2} \right) \sqrt{\vert g \vert} ~ d^{d}x
\end{equation}
on a region $ \Omega $ of a $ d $-dimensional manifold $ (\mathcal{M},g) $ takes the lattice form
\begin{equation}\label{slatt}
	S = \frac{1}{2} \sum_{\text{edges } ij} \frac{\mathcal{V}_{ij}}{\mathcal{l}_{ij}^{2}} ~ (\varphi_{i}-\varphi_{j})^{2} + \frac{1}{2} \sum_{\text{vertices } i} V_{i} ~ m^{2} \varphi_{i}^{2}
\end{equation}
Note that the field is defined on the vertices of the lattice. Here, $ \mathcal{l}_{ij}^{2} $ is is the \textit{proper edge length} and $ \mathcal{V}_{ij}, V_{i} $ stand for the \textit{dual edge} and \textit{vertex volumes}, respectively. The sums run over edges and vertices belonging to the lattice region corresponding to $ \Omega $. In case of a lattice composed of standardized $ d $-simplices, the proper edge length is given by two constants (one for spacelike edges, another for timelike edges) and the dual volumes are computed easily by counting distinct types of neighboring simplices.

Some authors have addressed the problem of coupling matter fields to quantum gravity \cite{Hamber1993, Paunkovic2016}; however, it is quite demanding both in theory and practice. In fact, even if we keep the geometry fixed, performing a consistent analysis of the field dynamics is not as simple as one could expect. Before we conclude this letter, we devote the last section to a brief outline of one particular framework serving this purpose.

\section{Discrete Evolution and Matter Fields}

If one views the Regge triangulation as a lattice instead as a piecewise-linear Lorentzian manifold, one obtains a system with a discrete notion of time, whose dynamics can be described by \textit{discrete canonical evolution} \cite{Dittrich2011, Dittrich2012, Dittrich2013}. It comes in two versions: a global one, which applies only if the lattice admits a foliation into non-intersecting time-slices (in very much the same way it is assumed within causal dynamical triangulations), and a local one, which applies to more general configurations. For simplicity, we shall discuss the global version. It starts with the lattice action
\begin{equation}\label{action}
	S = \sum_{n=0}^{t-1} S_{n+1}(x_{n},x_{n+1})
\end{equation}
composed of individual time-step contributions $  S_{n+1}(x_{n},x_{n+1}) $ and passes to the canonical picture upon defining momenta and obtaining a specific form of the equations of motion. Remarkably, the model allows for changing the number of degrees of freedom (i.e., the dimension of the configuration space $ \mathcal{Q}_{n} \ni x_{n} $) along the evolution, e.g. when the lattice expands or shrinks from one time-slice to the next. This irregularity gives rise to \textit{constraints} as well as \textit{free parameters} which need to be taken in account and have significant impact on the overall dynamics \cite{Dittrich2013}.

The analogical formalism for quantum systems was introduced in \cite{Hoehn2014a, Hoehn2014b} and subsequently applied to systems with quadratic action, which are known to posses linear equations of motion and can be therefore treated most easily \cite{Hoehn2014}. Discrete quantum evolution for systems with quadratic action was further investigated in \cite{Kaninsky2020} where it was argued that it is most naturally described by a non-unitary evolution map. The non-unitarity has its roots in the irregularity of the classical system, and craves for regularization (and occasionally renormalization) of final states. The theory comes with one-step propagators of the form
\begin{equation}\label{prop}
	_{\mathtt{c}} \langle \beta_{n+1} \vert \mathbb{U}_{n+1} \vert \gamma_{n} \rangle_{\mathtt{c}} = \mathcal{l}(V_{2}^{T} \beta_{n+1}) ~ \abs{\det \Sigma_{r} }^{1/2} ~ (2\pi)^{-q/2} ~ e^{iS_{n+1}(\gamma_{n},\beta_{n+1})}
\end{equation}
where $ \mathcal{l}(V_{2}^{T} \beta_{n+1}) $ is a regularization term and $ \abs{\det \Sigma_{r} }^{1/2} $ is a constant factor. The main part of the propagator is formed by the standard complex exponential featuring the one-step action contribution $ S_{n+1}(\gamma_{n},\beta_{n+1}) $ familiar from \eqref{action}. Upon the composition of multiple time-steps, the propagators \eqref{prop} make up the path integral.

Although the formalism of discrete canonical evolution was originally introduced in order to describe the dynamics of the lattice---or to say, the geometry---itself \cite{Dittrich2011}, it is suitable for various other systems: for instance, it can be easily applied to the case of scalar field on a fixed Lorentzian Regge lattice, as shown in \cite{Kaninsky2020}. What more, if one refrains from introducing higher order interaction terms, one can keep the action quadratic and benefit from a rather straightforward employment of the linear formalism. This way, it is possible to study the dynamical response of various quantum fields to a chosen geometry, very much like in quantum field theory on curved spacetime. As far as the author is concerned, this has not been done before. Within this letter, we would like to advocate for the application to a triangulated black hole spacetime. Note that such application could greatly benefit from the model of Dittrich and Loll \cite{Dittrich2006} which has two advantages: the standardized simplices make the lattice structure (and consequently the action) quite simple and the assumption that the lattice is foliated into separate time-steps allows one to use the global version of discrete canonical evolution. On the other hand, upon limiting the assumption of global foliation to a selected region, one can build similar models for exact spacetimes other than the Kottler solution \eqref{kottler}.

\begin{figure}[H]
	\centering
	\begin{tikzpicture}[scale=1]
		\tikzset{
			vertex/.style={
				shape=circle,fill=lightgray!100,minimum size=2mm,inner sep=0.2mm, label={[fill=none,label distance=1mm]90:#1}
			},
			edge/.style={
				draw,-,color=lightgray!100,line width=0.3mm
			},
			edget/.style={
				draw,dashed,color=lightgray!100,line width=0.3mm
			},
		}
		
		\coordinate (cia) at (-2,0);
		\coordinate (c1) at (-1,0);
		\coordinate (c2) at (0,0);
		\coordinate (c3) at (1,0);
		\coordinate (cib) at (2,0);
		
		\coordinate (cic) at (-1.5,0.866);
		\coordinate (c4) at (-0.5,0.866);
		\coordinate (c5) at (0.5,0.866);
		\coordinate (c6) at (1.5,0.866);
		\coordinate (cid) at (2.5,0.866);
		
		\coordinate (cie) at (-2,1.732);
		\coordinate (c7) at (-1,1.732);
		\coordinate (c8) at (0,1.732);
		\coordinate (c9) at (1,1.732);
		\coordinate (cif) at (2,1.732);	
		
		\coordinate (cig) at (-1.5,2.598);
		\coordinate (c10) at (-0.5,2.598);
		\coordinate (c11) at (0.5,2.598);
		\coordinate (c12) at (1.5,2.598);
		\coordinate (cih) at (2.5,2.598);

		\draw[edge] (c1) -- (c2) --(c3) -- (cib);
		\draw[edget] (c1) -- (c4) --(c2) -- (c5) -- (c3) -- (c6) -- (cib);
		\draw[edge] (c4) -- (c5) -- (c6);
		\draw[edget] (c4) -- (c8) -- (c5) -- (c9) -- (c6);
		\draw[edge] (c8) -- (c9);
		\draw[edget] (c8) -- (c11) -- (c9);

		\node[vertex] at (c1) {};
		\node[vertex] at (c2) {};
		\node[vertex] at (c3) {};
		\node[vertex] at (cib) {};
		
		\node[vertex] at (c4) {};
		\node[vertex] at (c5) {};
		\node[vertex] at (c6) {};
		
		\node[vertex] at (c8) {};
		\node[vertex] at (c9) {};
		
		\node[vertex] at (c11) {};

		\node[] at (3.5,0) {$ n = 0 $};
		\node[] at (3.5,0.866) {$ n = 1 $};
		\node[] at (3.5,1.732) {$ n = 2 $};
		\node[] at (3.5,2.598) {$ n = 3 $};
		
	\end{tikzpicture}
	\vspace{0 mm}
	\caption{A diagrammatic depiction of a shrinking lattice ending in a singularity. Spacelike edges are drawn in solid line, timelike edges are drawn in dashed line. Every time-step has one vertex less then the preceding one, which will result in a gradual information loss. Ultimately, the remaining information is disposed when it reaches the singularity at $ n = 3 $, which represents a spacelike boundary.}
	\label{fig:s}
	\vspace{4 mm}
\end{figure}
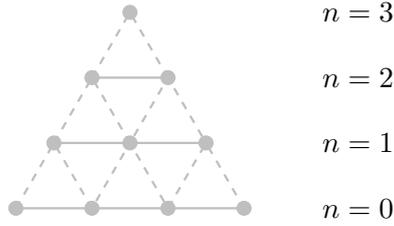

It can be expected that the non-unitary nature of the discrete quantum evolution will play a role in the triangulated black hole spacetime, which may result in information loss. Imagine the central singularity, which disposes all the information carried by the field once reached by it. In discrete spacetimes, such behavior is not limited to the spacetime boundary: information loss may occur at essentially any point of the evolution, since it depends on the numbers of neighboring vertices and their connectivity \cite{Hoehn2014, Kaninsky2020}. The idea is illustrated in Fig. \ref{fig:s}. The non-unitarity can be merely an artifact of discretization, but it can be also physical (like in the case of spacetime singularity). This distinction has some implications when it comes to the interpretation of final states \cite{Kaninsky2020}. The formalism is not free from complications, but it is flexible and suited for an immediate implementation.

\section{Conclusion}

Black holes are incredibly interesting gravitational phenomena attracting, among other things, constant attention of researchers from various fields. This still holds true in context of discrete spacetime. We have identified three main directions featuring discrete black hole models: numerical relativity and cosmology, effective models investigating the phenomenological implications of minimal length-scale, and discrete spacetime models in service of quantum gravity. We kept our focus on the latter family, which we find particularly remarkable, and looked closer at three of its successful representatives: causal sets, Regge triangulations, and causal dynamical triangulations. We found that in some way, all of them have been applied to black holes; the most important examples were briefly discussed. With small hyperbole, one could say that if a discrete spacetime model can describe a black hole, it is worth attention.

In the last section, we outlined a way towards inhabiting the discrete spacetime with quantum fields. We have limited our discussion to triangulations because they are typically easier to work with than causal sets. It is suggested that the quantum field on a fixed Lorentzian lattice could be described within the framework of discrete canonical evolution, especially if its action is only quadratic. The application to black holes is not only possible but also desirable, since it can directly address a number of questions concerning black hole phenomenology, either in connection to minimal length-scale or simply as a way to facilitate for a numerical simulation. We therefore encourage any interested researcher to discretize his or her own favorite black hole solution and provide it with some quantum matter: the tools are out there.

\section*{Acknowledgments}
This work was supported by Charles University Grant Agency [Project No. 906419].

\bibliographystyle{unsrt}
\renewcommand{\bibname}{Bibliography}
\bibliography{bibliography}

\begin{thebibliography}{10}

\bibitem{Buonanno2014}
Alessandra Buonanno and Bangalore Sathyaprakash.
\newblock Sources of gravitational waves: Theory and observations.
\newblock 2014.
\newblock arXiv:1410.7832.

\bibitem{Lousto2008}
Carlos Lousto and Yosef Zlochower.
\newblock Foundations of multiple-black-hole evolutions.
\newblock {\em Physical Review D}, 77, 2008.
\newblock arXiv:0711.1165.

\bibitem{Galaviz2010}
Pablo Galaviz, Bernd Bruegmann, and Zhoujian Cao.
\newblock Numerical evolution of multiple black holes with accurate initial
  data.
\newblock {\em Physical Review D}, 2010.
\newblock arXiv:1004.1353.

\bibitem{Sperhake2013}
U.~Sperhake.
\newblock Black holes on supercomputers: Numerical relativity applications to
  astrophysics and high-energy physics.
\newblock {\em Acta Phys. Polon. B}, 44(12):2463--2536, 2013.

\bibitem{Okawa2014}
Hirotada Okawa, Helvi Witek, and Vitor Cardoso.
\newblock Black holes and fundamental fields in numerical relativity: initial
  data construction and evolution of bound states.
\newblock {\em Physical Review D}, 89, 2014.

\bibitem{Fendt2019}
C.~Fendt.
\newblock Approaching the black hole by numerical simulations.
\newblock {\em Universe}, 2019.
\newblock arXiv:1907.12789.

\bibitem{Bentivegna2018}
Eloisa Bentivegna, Timothy Clifton, Jessie Durk, Miko\l{}aj Korzy\'nski, and
  Kjell Rosquist.
\newblock Black-hole lattices as cosmological models.
\newblock {\em Class. Quant. Grav.}, 35(17), 2018.
\newblock arXiv:1801.01083.

\bibitem{Gregoris2020}
D~Gregoris and K.~Rosquist.
\newblock Observational backreaction in discrete black holes lattice
  cosmological models.
\newblock 35(14), 2020.
\newblock arXiv:2006.00855.

\bibitem{Garay1994}
Luis Garay.
\newblock Quantum gravity and minimum length.
\newblock {\em International Journal of Modern Physics A}, 10, 1994.

\bibitem{Hossenfelder2013}
Sabine Hossenfelder.
\newblock Minimal length scale scenarios for quantum gravity.
\newblock {\em Living Rev. Rel.}, 2013.
\newblock arXiv:1203.6191.

\bibitem{Padmanabhan2015}
T.~Padmanabhan.
\newblock Distribution function of the atoms of spacetime and the nature of
  gravity.
\newblock {\em Entropy}, 17:7420--7452, 2015.
\newblock arXiv:1508.06286.

\bibitem{Hooft2016}
Gerard ’t Hooft.
\newblock How quantization of gravity leads to a discrete space-time.
\newblock {\em J. Phys.: Conf. Ser.}, 701(012014), 2016.

\bibitem{Szabo2002}
R.J Szabo.
\newblock {BUSSTEPP} lectures on {s}tring theory: An introduction to {s}tring
  theory and {D}-brane dynamics.
\newblock 2002.
\newblock arXiv:hep-th/0207142.

\bibitem{Gambini2011}
Rodolfo Gambini and Jorge Pullin.
\newblock {\em A first course in loop quantum gravity}.
\newblock Oxford University Press, 2011.

\bibitem{Henson2010}
Joe Henson.
\newblock Discovering the discrete universe.
\newblock {\em Foundations of Space and Time: Reflections on Quantum Gravity},
  2010.
\newblock arXiv:1003.5890.

\bibitem{Mead1966}
C.~Alden Mead.
\newblock Observable consequences of fundamental-length hypotheses.
\newblock {\em Phys. Rev.}, 143(990), 1966.

\bibitem{Maziashvili2013}
Michael Maziashvili.
\newblock Minimum length, extra dimensions, modified gravity and black hole
  remnants.
\newblock {\em Journal of Cosmology and Astroparticle Physics},
  2013(03):042--042, 2013.

\bibitem{Ali2012}
Ahmed~Farag Ali.
\newblock No existence of black holes at {LHC} due to minimal length in quantum
  gravity.
\newblock {\em JHEP}, 09, 2012.
\newblock arXiv:1208.6584.

\bibitem{Bambi2008}
Cosimo Bambi and Katherine Freese.
\newblock Dangerous implications of a minimum length in quantum gravity.
\newblock {\em Class. Quant. Grav.}, 25, 2008.

\bibitem{Corley1998}
Steven Corley and Ted Jacobson.
\newblock Lattice black holes.
\newblock {\em Phys. Rev. D}, 57:6269--6279, 1998.
\newblock arXiv:hep-th/9709166.

\bibitem{Mejrhit2019}
Karim Mejrhit.
\newblock Non-extensive entropy and discrete quantum spectrum of black holes
  from quantum geometry.
\newblock {\em Physics Letters B}, 775:32--36, 2019.

\bibitem{Lowe2015}
David~A. Lowe and Larus Thorlacius.
\newblock Quantum information erasure inside black holes.
\newblock {\em JHEP}, 12:096, 2015.
\newblock arXiv:1508.06572.

\bibitem{Spallucci2011}
Euro Spallucci and S.~Ansoldi.
\newblock Regular black holes in {UV} self-complete quantum gravity.
\newblock {\em Physics Letters B}, 701, 2011.

\bibitem{Bekenstein1999}
J.~D. Bekenstein.
\newblock Quantum black holes as atoms.
\newblock {\em Prodeedings of the Eight Marcel Grossmann Meeting, World
  Scientific Singapore}, pages 92--111, 1999.
\newblock arXiv:gr-qc/9710076.

\bibitem{Corda2020}
Christian Corda and Fabiano Feleppa.
\newblock The quantum black hole as a gravitational hydrogen atom.
\newblock 2020.
\newblock arXiv:1912.06478.

\bibitem{Surya2019}
Sumati Surya.
\newblock The causal set approach to quantum gravity.
\newblock {\em Living Reviews in Relativity}, 22, 2019.

\bibitem{Hamber2009}
Herbert~W. Hamber.
\newblock {\em Quantum Gravitation – The Feynman Path Integral Approach}.
\newblock Springer-Verlag Berlin Heidelberg, 2009.

\bibitem{Barrett2018}
John~W. Barrett, Daniele Oriti, and Ruth~M. Williams.
\newblock Tullio {R}egge's legacy: {R}egge calculus and discrete gravity.
\newblock 2018.
\newblock arXiv:1812.06193.

\bibitem{Loll2019}
R.~Loll.
\newblock Quantum gravity from causal dynamical triangulations: A review.
\newblock {\em Class. Quant. Grav.}, 37(1), 2019.
\newblock arXiv:1905.08669.

\bibitem{Dou2003}
D.~Dou and R.D. Sorkin.
\newblock Black-hole entropy as causal links.
\newblock {\em Foundations of Physics}, 33:279–296, 2003.
\newblock arXiv:gr-qc/0302009.

\bibitem{Rideout2006}
D.~Rideout and S.~Zohren.
\newblock Evidence for an entropy bound from fundamentally discrete gravity.
\newblock {\em Class. Quant. Grav.}, 23:6195--6213, 2006.
\newblock arXiv:gr-qc/0606065.

\bibitem{He2008}
Song He and David Rideout.
\newblock A causal set black hole.
\newblock {\em Class. Quant. Grav.}, 26(12), 2008.
\newblock arXiv:0811.4235.

\bibitem{Asato2019}
Yu~Asato.
\newblock Black holes and singularities in causal set gravity.
\newblock {\em Class. Quant. Grav.}, 36(195008), 2019.
\newblock arXiv:1905.03827.

\bibitem{Surya2020}
Sumati Surya, Nomaan X, and Yasaman~K. Yazdi.
\newblock Entanglement entropy of causal set de {S}itter horizons.
\newblock 2020.
\newblock arXiv:2008.07697.

\bibitem{Wong1971}
Cheuk‐Yin Wong.
\newblock Application of {R}egge calculus to the {S}chwarzschild and
  {R}eissner‐{N}ordstr\"{o}m geometries at the moment of time symmetry.
\newblock {\em Journal of Mathematical Physics}, 12(70), 1971.

\bibitem{Khatsymovsky2020}
V.M. Khatsymovsky.
\newblock On the discrete version of the {S}chwarzschild problem.
\newblock {\em Universe}, 6(10):185, 2020.
\newblock arXiv:2008.13756.

\bibitem{Dittrich2006}
B.~Dittrich and R.~Loll.
\newblock Counting a black hole in {L}orentzian product triangulations.
\newblock {\em Class. Quant. Grav.}, 23:3849--3878, 2006.
\newblock arXiv:gr-qc/0506035.

\bibitem{Sorkin1975}
Rafael Sorkin.
\newblock The electromagnetic field on a simplicial net.
\newblock {\em Journal of Mathematical Physics}, 16(2432), 1975.

\bibitem{Hamber1993}
Herbert~W. Hamber and Ruth~M. Williams.
\newblock Simplicial gravity coupled to scalar matter.
\newblock {\em Nucl. Phys. B}, 415:463--496, 1993.
\newblock arXiv:hep-th/9308099.

\bibitem{McDonald2010a}
Jonathan~R. McDonald and Warner~A. Miller.
\newblock Coupling non-gravitational fields with simplicial spacetimes.
\newblock {\em Class. Quant. Grav.}, 27:095011, 2010.
\newblock arXiv:1002.5001.

\bibitem{Paunkovic2016}
Nikola Paunkovic and Marko Vojinovic.
\newblock Gravity-matter entanglement in {R}egge quantum gravity.
\newblock {\em J. Phys. Conf. Ser.}, 701(1), 2016.
\newblock arXiv:1601.06831.

\bibitem{Sverdlov2008}
Roman Sverdlov and Luca Bombelli.
\newblock Gravity and matter in causal set theory.
\newblock {\em Class. Quant. Grav.}, 26, 2008.
\newblock arXiv:0801.0240.

\bibitem{Johnston2008}
Steven Johnston.
\newblock Particle propagators on discrete spacetime.
\newblock {\em Class. Quant. Grav.}, 25:202001, 2008.
\newblock arXiv:0806.3083.

\bibitem{Johnston2010}
Steven Johnston.
\newblock Quantum fields on causal sets.
\newblock 2010.
\newblock arXiv:1010.5514.

\bibitem{DableHeath2020}
Edmund Dable-Heath, Christopher Fewster, Kasia Rejzner, and Nick Woods.
\newblock Algebraic classical and quantum field theory on causal sets.
\newblock {\em Class. Quant. Grav.}, 2020.
\newblock arXiv:1908.01973.

\bibitem{Joshi1993}
P.S. Joshi.
\newblock {\em Global aspects in gravitation and cosmology}.
\newblock Number~87 in International series of monographs on physics. Oxford
  Universily Press, 1993.

\bibitem{Eichhorn2019}
Astrid Eichhorn, Sumati Surya, and Fleur Versteegen.
\newblock Induced spatial geometry from causal structure.
\newblock {\em Class. Quant. Grav.}, 2019.
\newblock arXiv:1809.06192.

\bibitem{Regge1961}
T.~Regge.
\newblock General relativity without coordinates.
\newblock {\em Nuovo Cimento}, 19:558--71, 1961.

\bibitem{Ambjorn2005}
J.~Ambj{\o}rn, B.~Durhuus, and T.~Johnsson.
\newblock {\em Quantum Geometry: A statistical field theory approach}.
\newblock Cambridge Monographs on Mathematical Physics. Cambridge University
  Press, 2005.

\bibitem{Cheeger1984}
Jeff Cheeger, Werner Muller, and Robert Schrade.
\newblock On the curvature of piecewise flat spaces.
\newblock {\em Communications in Mathematical Physic}, 1984.

\bibitem{Oriti2009}
D.~Oriti, editor.
\newblock {\em Approaches to Quantum Gravity: Toward a New Understanding of
  Space, Time and Matter}.
\newblock Cambridge University Press, 2009.

\bibitem{Romer1986}
H.~R\"omer and M.~Z\"ahringer.
\newblock Functional integration and the diffeomorphism group in {E}uclidean
  lattice quantum gravity.
\newblock {\em Class. Quantum Grav.}, 3:897--910, 1986.

\bibitem{Ambjoern2013}
J.~Ambjørn, A.~Gorlich, J.~Jurkiewicz, and R.~Loll.
\newblock Quantum gravity via causal dynamical triangulations.
\newblock 2013.
\newblock arXiv:1302.2173.

\bibitem{Dittrich2011}
Bianca Dittrich and Philipp~A. Höhn.
\newblock From covariant to canonical formulations of discrete gravity.
\newblock {\em Class. Quant. Grav.}, 27(15), 2011.
\newblock arXiv:0912.1817.

\bibitem{Dittrich2012}
Bianca Dittrich and Philipp~A. Höhn.
\newblock Canonical simplicial gravity.
\newblock {\em Class. Quant. Grav.}, 29(11), 2012.
\newblock arXiv:1108.1974.

\bibitem{Dittrich2013}
Bianca Dittrich and Philipp~A. Höhn.
\newblock Constraint analysis for variational discrete systems.
\newblock {\em Journal of Mathematical Physics}, 54:093505, 2013.
\newblock arXiv:1303.4294.

\bibitem{Hoehn2014a}
Philipp~A. Höhn.
\newblock Quantization of systems with temporally varying discretization i:
  Evolving {H}ilbert spaces.
\newblock {\em Journal of Mathematical Physic}, 55(083508), 2014.
\newblock arXiv:1401.6062.

\bibitem{Hoehn2014b}
Philipp~A. Höhn.
\newblock Quantization of systems with temporally varying discretization. ii:
  Local evolution moves.
\newblock {\em Journal of Mathematical Physics}, 55(103507), 2014.
\newblock arXiv:1401.7731.

\bibitem{Hoehn2014}
Philipp~A. Höhn.
\newblock Classification of constraints and degrees of freedom for quadratic
  discrete actions.
\newblock {\em Journal of Mathematical Physics}, 55(113506), 2014.

\bibitem{Kaninsky2020}
Jakub Káninský.
\newblock Models of discrete linear evolution for quantum systems.
\newblock 2020.
\newblock arXiv:2011.08715.

\end{thebibliography}

\end{document}